\documentclass[conference]{IEEEtran}
\IEEEoverridecommandlockouts

\usepackage{amsmath,amssymb,amsfonts}
\usepackage{subfig}

\usepackage{xspace}
\usepackage{paralist}
\usepackage{hyperref}
\usepackage{cleveref}

\usepackage{booktabs}
\usepackage{multirow}
\usepackage{graphicx}
\usepackage{xcolor}
\usepackage{array}

\newcolumntype{R}[2]{%
    >{\adjustbox{angle=#1,lap=\width-(#2)}\bgroup}%
    l%
    <{\egroup}%
}

\newcommand{\blockx}{\textsc{BlockChain I/O}\xspace}
\newcommand{\pie}{\textsc{PieChain}\xspace}
\newcommand{\croc}{\textsc{CroCoDai}\xspace}

\begin{document}

\title{\blockx : Enabling Cross-Chain Commerce
}
\author{\IEEEauthorblockN{Anwitaman Datta, Dani\"el Reijsbergen, Jingchi Zhang, and Suman Majumder} \IEEEauthorblockA{{Nanyang Technological University} \\
Singapore, Singapore\\
\text{\{anwitaman,daniel.reijsbergen,suman.majumder\}@ntu.edu.sg} \\
\text{jingchi001@e.ntu.edu.sg}}}


%



\maketitle

\begin{abstract}
  Blockchain technology enables secure tokens transfers in digital marketplaces, and recent advances in this field provide other desirable properties such as  efficiency, privacy, and price stability. However, these properties do not always generalize to a setting across \textit{multiple} independent blockchains. Despite the growing number of existing blockchain platforms, there is a lack of an overarching framework whose components provide all of the necessary properties for practical cross-chain commerce. We present \blockx to provide such a framework. \blockx introduces entities called \textit{cross-chain services} to relay information between different blockchains. The proposed design ensures that cross-chain services cannot violate transaction safety, and they are furthermore disincentivized from other types of misbehavior through an audit system. \blockx uses native stablecoins to mitigate price fluctuations, and a decentralized ID system to allow users to prove aspects of their identity without violating privacy. After presenting the core architecture of \blockx, we demonstrate how to use it to implement a \textit{cross-chain marketplace} and discuss how its desirable properties continue to hold in the end-to-end system. Finally, we use experimental evaluations to demonstrate \blockx's practical performance.
\end{abstract}

\section{Introduction}
\label{sec:introduction}
Numerous blockchains, distributed ledgers,\footnote{For the remainder of this paper, we use the term `blockchain' loosely to also include other kinds of distributed ledgers even though blockchains are technically a subcategory, e.g., Avalanche's Directed Acyclic Graphs (DAGs) belong to the latter category but not the former \cite{rocket2019scalable}.} and Decentralized Finance (DeFi) products have been designed and deployed, resulting in siloed ecosystems since most of these systems do not support interoperation with others by default. 
To ameliorate the limitations of isolated blockchains, several works have aimed at enabling different extents of intercommunication, interoperation, and integration across blockchains. These include theoretical building blocks as well as functional software artifacts, with some that are limited to specific subsets of systems, whereas others are more general-purpose \cite{zamyatin2021sok,gudgeon2019sok,belchior2021survey,dltsurvey2023}. 
Among these, the most basic are generic payload-agnostic inter-blockchain communication mechanisms \cite{goes2020interblockchain,herlihy2022cross,HerlihyFaultTolerantCCS} and protocols for value transfer, payments, or exchanges across blockchains \cite{interledger2015protocol,crosschainbvirtualpayments,universalatomicswap} which assume and enable upper-layer applications to orchestrate the business logic.
Other examples include dedicated bridges that validate transfers between a given pair of blockchains \cite{bridges,xie2022zkbridge}, mechanisms that ensure the security of cross-chain transfers through timelocks \cite{herlihy2018atomic,universalatomicswap}, and relay chains that support efficient state proofs \cite{micali2021compact,bonneau2020mina}.

\begin{figure}[tp!]
\centering
\includegraphics[width=\linewidth]{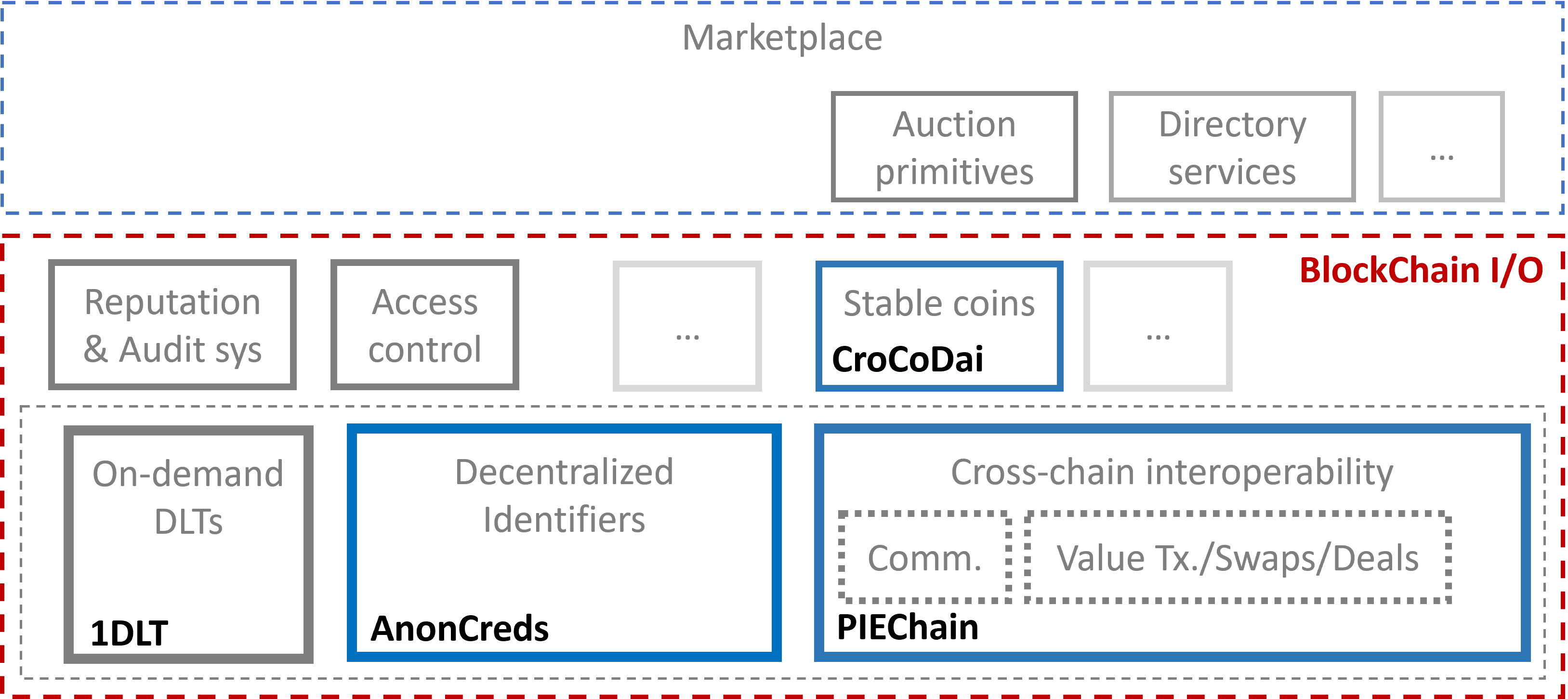}
\caption{The \blockx stack for a versatile cross-chain platform with an overlying open marketplace: This paper primarily focuses on the modules highlighted in \textcolor{blue}{blue}.}
\label{fig:blockxarch}
\end{figure}




The wealth of proposals mentioned above indicates that there is a rich set of existing solutions that ensure the basic security of cross-chain interactions, i.e., that tokens are never stolen (safety) and that transfers are never stalled forever (liveness). However, as in Maslow's hierarchy of needs, there are desirable properties \textit{beyond} basic security in the context of Web 3.0: examples include privacy, fairness, token price stability, and efficiency while maintaining support for a wide range of use cases.
Although approaches exist that address some of these aspects in isolation, such as privacy \cite{deshpande2020privacy}, stablecoin support \cite{crocodai}, and generality \cite{piechain}, there is a need for an umbrella framework that achieves all of the desired primitives while ensuring that no properties provided by one component are violated by another component, and that all potential synergies between the components are exploited. 
Recent government initiatives that propose or include online marketplaces, e.g., the Indian Government's Open Network for Digital Commerce (ONDC \cite{ONDC}) and the EU's strategy paper on virtual worlds \cite{metaverseEU}, are useful starting points both for the aspired nature of open marketplaces in general and also in part in determining the list of desired primitives for cross-chain commerce. In this work, we define cross-chain commerce as any large-scale exchange of digital tokens on multiple independent blockchains, and any platform that facilitates such exchanges to be a cross-chain marketplace.



To that end, we enumerate a set of desirable functionalities required as building blocks for versatile cross-chain commerce, and describe a modular technology stack, \blockx (see \Cref{fig:blockxarch}) that achieves these desiderata. While not exhaustive by any means, this set comprises common and crucial functions we gleaned from the literature describing versatile marketplaces, and provides a reasonable baseline; future works could identify and build on this to add further functions.
To ensure safety in a cross-chain setting, \blockx uses \pie:\footnote{Presented as a demo, the original \pie work \cite{piechain} focused on the UI \& example uses of the system. This work, which, among other things, extends the previous \pie work, focuses on the technical description and practical evaluation.} dedicated entities called \textit{cross-chain services} (CC-SVCs) relay relevant information, e.g., bids, asset prices, and identity information, between nodes. Once the outline of a deal has become clear, e.g., after an auction has terminated, a CC-SVC sends a tentative exchange of tokens to the involved blockchains, and the users lock their tokens in escrow. Users then vote to \textit{commit} if the exchange of tokens is agreeable, and abort otherwise. As such, \blockx facilitates \textit{cross-chain deals} \cite{herlihy2022cross}, which have proven security guarantees. If a CC-SVC violates liveness by going offline while processing a deal, then this is provable to nodes who monitor the relevant blockchains. \blockx utilizes a separate class of nodes called \textit{auditors} to detect this misbehavior and relay it to a governance layer -- the resulting reputation damage provides an incentive for CC-SVCs to behave honestly. In \blockx, the same auditors who detect liveness violations are used to detect fairness violations, allowing for the same reputation infrastructure to be leveraged for multiple purposes. To provide decentralized identities that preserve privacy, we utilize Hyperledger AnonCreds \cite{anoncreds}. For a cross-chain stablecoin, we leverage \croc \cite{crocodai}, which can be implemented for cross-chain commerce and integrated with the core interoperability module, \pie.

To illustrate \blockx's generality, we present three use cases -- a cross-chain marketplace, scalping-resistant ticket sales, and Sybil-resistant reputations -- for cross-chain commerce whose challenges are overcome by the framework's components. Furthermore, to validate the implementation of our ideas and viability of a versatile platform for cross-chain commerce, we present a proof-of-concept implementation of the first use case, i.e., a decentralized marketplace that allows users to create and bid on token listings. This implementation provides a proof-of-concept and drives our performance benchmark experiments. To aid reproducibility, the code for our implementation has been made public at \url{https://github.com/ntublockchain/I-O}.


In summary, our contributions are as follows.
\begin{itemize}
    \item We survey recent proposals for open marketplaces, e.g., the Indian government's ONDC \cite{ONDC} and the EU's strategy paper on virtual worlds \cite{metaverseEU}, and present a three-tier stack of desirable properties for cross-chain commerce \Cref{sec:initiatives}. {This property stack may provide a starting point for future work on cross-chain commerce.}
    \item We present \blockx, a framework for cross-chain commerce (\Cref{sec:architecture}). We describe the main components, including \pie and \croc, and their integration.
     We also describe in \Cref{sec:use_cases} three use cases in which multiple of these properties must be satisfied simultaneously.
    \item We present a proof-of-concept implementation of a decentralized marketplace that is built using \blockx (\Cref{sec:decentralized_marketplace}).  {As our code is publicly available, future implementations of cross-chain marketplace may reuse part of our work, e.g., the code for CC-SVCs, to facilitate deployment. Furthermore, we demonstrate that our implementation has practical performance (\Cref{sec:experiments}), as it is able to support thousands of bids on concurrent auctions.}
\end{itemize} 
The rest of this paper is organized as follows. \Cref{sec:background} presents the background and related work. \Cref{sec:initiatives} surveys existing proposals for open marketplaces to motivate a list of desiderata for cross-chain commerce. \Cref{sec:architecture} presents the core architecture of \blockx. \Cref{sec:use_cases} discusses three cross-chain commerce use cases demonstrating the need for \blockx's properties. \Cref{sec:decentralized_marketplace} presents a decentralized marketplace built on top of \blockx.
\Cref{sec:experiments} presents the performance benchmark results and \Cref{sec:conclusions} concludes the paper.

\section{Background \& Related Work}
\label{sec:background}
The aim of this section is to provide background information on several core components of cross-chain commerce -- blockchain interoperability, digital identities, and reputation systems -- through a discussion of academic related work and existing blockchain platforms. We will also highlight the differences between \blockx and related blockchain interoperability frameworks, e.g., Polkadot and Cosmos. We assume that readers are familiar with blockchain technology basics such as transactions and smart contracts \cite{narayanan2016bitcoin}.

\subsection{Blockchain Interoperability}
Blockchains transactions are atomic by design, i.e., if a single transaction consists of multiple steps, then if any step is committed/aborted, then all are. However, atomicity cannot be guaranteed by default in \textit{cross-chain systems} where transactions may involve steps on different blockchains, which has led to the emergence of dedicated blockchain interoperability solutions.
Existing interoperability solutions can be typically categorized as sidechains, relays, notary schemes, or ledgers of ledgers \cite{belchior2021survey}.

A \textit{sidechain} is a blockchain that interacts with another (typically primary) blockchain as an extension, aiming to improve its scalability or interoperability. Major examples of sidechains in the context of Bitcoin include RSK \cite{lerner2022rsk} and the Liquid Network \cite{nick2020liquid}.
A \textit{notary scheme} is a system where an entity initiates a transaction on one blockchain in response to a specific event occurring on another blockchain -- one example is the \pie framework that we use in the core architecture of \Cref{sec:architecture}.
Similarly, a \textit{relay} refers to a mechanism where a designated entity keeps track of events or transactions on one blockchain and then relays this information to another blockchain. One of the most popular relay solutions, BTC Relay \cite{btcrelay}, was released in 2016 by the Ethereum Foundation. 
Finally, a `\textit{ledger of ledgers}' system is one where a central blockchain is connected to multiple other blockchains (known as sidechains or parachains). These blockchains collectively form an interconnected ecosystem. 

Polkadot \cite{polkdot2020} is a prominent example of a ledger of ledgers. It relies on an underlying relay chain for security, which can be used by other parachains (parallel chains), which could in principle run distinct protocols, inducing in effect a logical star topology with the relay chain in the center.  Thus Polkadot achieves not only sharding to improve scalability, but also provides support for heterogeneity, and given that all the parachains rely on the same relay chain, the parachains can natively interoperate. However, this does not immediately help in solving the larger problem of facilitating arbitrary existing blockchain pairs which do not share Polkadot's relay chain, which is addressed in \blockx{}. 
Similar to Polkadot, Cosmos \cite{cosmos20} also uses a central `hub' which ensures governance at a global level, while supporting parallel chains called `zones'. 

\subsection{Digital Identities}
A \textit{digital identity} connects an individual's \textit{attributes}, e.g., her name, age, location, or reputation, to a digital presence such as an online account or public key.
Different methods exist for storing and sharing digital identities -- for example, they can be provided by a corporation (e.g., Google), by the individual herself, or by a public ledger. Digital identity systems that are decentralized -- i.e., not maintained by a single entity -- are also called \textit{Self-Sovereign Identity} (SSI) systems \cite{eer2022bottom}. Individual SSI data entries that contain attribute information are called \textit{Decentralized Identifiers} (DIDs), and a database that contains DIDs is commonly called a \textit{Verifiable Data Registry} (VDR). One prominent example of an SSI is the Sovrin Network \cite{windley2021sovrin} which uses a public blockchain as a VDR. However, a major challenge in SSI systems is establishing trust between identity issuers and validators \cite{eer2022bottom}: the validator must decide whether a DID and its attribute information come from a trusted source.
In Hyperledger AnonCreds \cite{anoncreds}, this challenge is addressed by assigning the creation and verification of digital identities a variety of user types, particularly \textit{holders} who have digital attribute information, \textit{issuers}  who issue DIDs, and \textit{verifiers} who verify DIDs.
AnonCreds specifies a set of protocols for zero-knowledge proofs and schema definitions that allow any consortium of users to run an SSI system on a permissioned blockchain, e.g., Hyperledger Indy, where the consortium members have write access and arbitrary users can have read access. AnonCreds is inherently decentralized -- subject to acceptance of participants within an ecosystem, arbitrary entities may participate as issuers, in the creation of VDRs, or the creation of DIDs given a VDR. SSI schemes provide privacy through the use of zero-knowledge cryptography to prove attributes from a  DID, and accountability because issuers sign the DIDs so that issuers of incorrect DIDs suffer reputation damage.

\subsection{E-Commerce \& Reputation Systems}
\label{sec:ecommerce}

E-commerce refers to the electronic sale of goods or services -- a prominent example is an online marketplace in which a website is maintained by a dedicated entity (e.g., Amazon or eBay), and a multitude of independent vendors create listings that allows customers to browse and bid on items.  Depending on the marketplace, vendors can set a fixed price for each item (e.g., Amazon), or buyers can bid for the items through an \textit{e-auction} mechanism (e.g., e-Bay). 
Recent advances in multi-party computation and zero-knowledge cryptography have enabled e-auction approaches that are both privacy-preserving and verifiable \cite{bag2019seal,blass2018strain}, thus enabling e-auctions on public blockchains \cite{galal2019verifiable}.


\textit{Trust} \cite{beatty2011consumer} is a critical factor that determines the success of e-commerce platforms.
Vendors can establish trust through repeated interactions with buyers, generating (if successful) positive feedback. In an online marketplace, vendor reputation metrics can be computed automatically from feedback and displayed alongside listings. For example, on eBay, the percentage of positive feedback is displayed on each vendor's account page. Privacy \cite{hasan2022privacy} is an important aspect of reputation systems: if user identities are known, then users may avoid giving negative feedback out of fear of retaliation. However, full anonymity may allow vendors to inflate their reputation (or damage their competitors') through dummy accounts -- an example of a so-called \textit{Sybil attack}. 
Recent advances in reputation systems include a blockchain-based e-commerce platform in which buy orders are pooled and sellers compete to fill the order \cite{martins2020fostering}: this raises the cost of Sybil attacks as fake buyer(s) who collude with sellers to boost the seller's rating risk being obligated to purchase a real item if an honest seller wins the auction. 
Finally, Beaver \cite{soska2016beaver} is a decentralized anonymous marketplace in which the cost of a Sybil attack can be made explicit.

\section{Cross-Chain Commerce Desiderata}
\label{sec:initiatives}
In this section, we discuss several recent government-led initiatives that sketch a vision for the functionalities of open marketplaces. These initiatives are more aspirational and take a broader view than the related work from \Cref{sec:background}, and allow to us identify and refine key desiderata for cross-chain commerce.
We focus on two key documents: the Indian government's ONDC \cite{ONDC} and the EU's strategy paper on virtual worlds \cite{metaverseEU} because they i.\ are recent (from 2022 and 2023, respectively), ii.\ consider the possibility of integrating multiple independent networks underpinned by blockchains, and iii.\ describe end-to-end functionalities for open markets. 
Other initiatives and strategy papers focus on specific aspects of digital markets (e.g., the UK's Digital Markets, Competition and Consumers Bill \cite{ukdmccbill} and the EU's Digital Market Act \cite{eudma}, which respectively focus on competition and gatekeepers), do not have detailed publicly available documentation (e.g., China's RealDID chain \cite{realDID}), or are more than 5 years old and/or do not consider interoperability (e.g., blockchain strategy papers by McKinsey \cite{mckinsey}, Deloitte \cite{deloitte}, and PWC \cite{pwc}). 
We also discuss existing systems that have the potential for integration with the cross-chain marketplace, such as central bank digital currencies (CBDCs) and digital identity systems.

\begin{figure}[tp!]
\centering
\includegraphics[width=\linewidth]{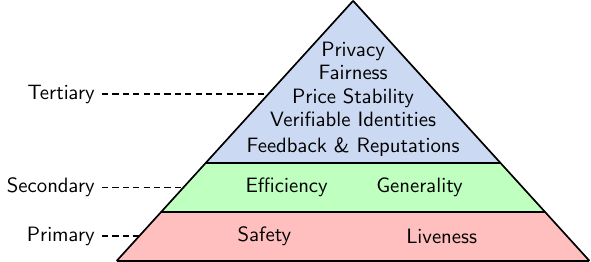}
\caption{A hierarchy of requirements for a decentralized marketplace.}
\label{fig:pyramid}
\end{figure}


\subsection{Cross-Chain Market Initiatives}

\subsubsection*{ONDC}
India's ONDC proposal presents a vision of a unified open marketplace in which shared protocols enable participants from different platforms to interact without barriers (Fig.\ 2 of \cite{ONDC}). Blockchains are mentioned as an enabler of  such platforms (Section 2.3.3), and many of the core features of ONDC's vision of an open network (Section 2.2.1) -- e.g., decentralized, interoperable, and with minimal governance -- are also key features of blockchains.
Fig.\ 7 in \cite{ONDC} visualizes the envisioned system, including an explicit mention of privacy-preserving data exchanges and the system's key components, i.e., the (account) registry, network policies, reputation ledger, open data ledger, payment processor, and inter-network interoperability. Although several of these are satisfied by default in a cross-chain setting (e.g., payment processing, a data ledger, and interoperability), the others -- \textit{privacy}, \textit{reputation}, and \textit{account registries} -- are higher-tier desiderata. Regarding privacy, the ONDC's definition (Section 3.4.2) is that transaction data is not stored on the global level and that consumer's personally identifiable information is not leaked. 

\subsubsection*{Virtual Worlds}

The EU's recent strategy paper on virtual worlds \cite{metaverseEU} contains a substantial section on digital markets (Section 3.3.1). The document's main text contains multiple references to blockchains, which are described as the core technology underlying the economy of virtual worlds. Blockchains are mentioned most frequently in the section on economic challenges for virtual worlds because of their reliability and tamper-proofness (Section 3.3.1). Scalability, interoperability, environmental footprint, security, and volatility vis-\'a-vis government-issued money are mentioned as drawbacks of blockchains, although it is mentioned that stablecoins mitigate the latter. Privacy is also extensively mentioned as a challenge for virtual worlds in general (i.e., beyond digital markets), as is authentication (Section 3.3.2). Whereas scalability and low environmental footprint are fundamental properties of any blockchain network, \textit{privacy}, \textit{authentication}, and \textit{price stability}.

\subsection{Potential Digital Market Components}

Several other government initiatives that have the potential for integration with open markets have recently been implemented or proposed. The first is that of unified \textit{digital identity systems} that enable citizen authentication. Example include India's Aadhaar, the EU's proposed EUDI, Australia's MyGov, and Singapore's SingPass. Currently, such systems typically rely on a centralized architecture maintained by a government institution, without interoperability beyond services supported by the platform. However, we envision that these architectures can be extended to provide time-bounded credentials that are uploaded to a blockchain, as such providing authentication for cross-chain commerce. 

\textit{Central bank digital currencies} (CBDCs) are a second type of component that is well-suited for integration with a cross-chain marketplace. Examples of (proposals for) such currencies include China's digital renminbi and the EU's digital euro. CBDCs have (much) less price volatility than typical cryptocurrencies, and have less risk of collapse \cite{briola2023anatomy} than regular stablecoins since they are guaranteed by a national government. Although such initiatives are not always implemented through blockchains, they promise to facilitate efficient exchanges of digital coins which makes them amenable to provide price stability for cross-chain commerce.


\subsection{Common Desiderata}

Based on the above, we divide the desired properties for cross-chain commerce into three tiers (see also \Cref{fig:pyramid} for a visualization). In particular, these are \textit{primary} properties that should hold in any blockchain system, \textit{secondary} properties that should hold in any blockchain application that works at scale, and the \textit{tertiary} properties that must hold for viable cross-chain commerce and which can be found in recent digital marketplace initiatives.


\subsubsection*{Primary Properties}
The first and most basic desirable properties of a cross-chain framework is that transactions are \textit{secure}. In the academic literature, security has typically been formalized through the properties of \textbf{safety} (i.e., bad things never happen) and \textbf{liveness} (i.e., good things eventually happen) \cite{cachin2001secure}. In cross-chain commerce, in which multiple parties agree to an exchange of tokens, the essential safety requirement is that token trades are \textit{atomic} -- i.e., all transfers in the trade are committed or none are. This ensures that users who send tokens also receive the tokens agreed upon in the deal, i.e., if the winning bidder in an e-commerce auction sends a payment then she also receives the listed asset from the seller and vice versa. 

\subsubsection*{Secondary Properties}
Once basic security is attained, the second tier of desirable properties 
are assumed to be a requirement of many but not all users -- although a system that fails to satisfy one of these properties may still have use for a niche of users, it would struggle to achieve mass adoption.
In our context, this requires that the system provides \textbf{efficiency} (i.e., transaction times or costs are not prohibitive), and \textbf{generality} (i.e., a wide range of use cases and ledger protocols are supported). Such properties have been studied extensively in the scientific literature (e.g., sharding for efficiency \cite{dang2019towards} and secure smart contract languages for generality \cite{nikolic2018finding}), and are crucial for any large-scale application, including cross-chain commerce.


\subsubsection*{Tertiary Properties}
Finally, to achieve mass adoption, a cross-chain platform must also provide properties that are requirements for large minorities of users.
From the initiatives earlier in this section, we can identify \textbf{verifiable identities} for user authentication, \textbf{privacy} through pseudonymity, a \textbf{reputation} system, and \textbf{price stability}. An additional tertiary property found in the academic literature is \textbf{fairness} \cite{dreier2013formal} -- i.e., although  basic safety guarantees that users cannot lose tokens without receiving an agreed compensation, the deal itself must have been established fairly, i.e., without scope for manipulation or bribery by other parties. 

\section{The \blockx Framework}
\label{sec:architecture}
\begin{figure}[tp!]
\centering
\includegraphics[width=\linewidth]{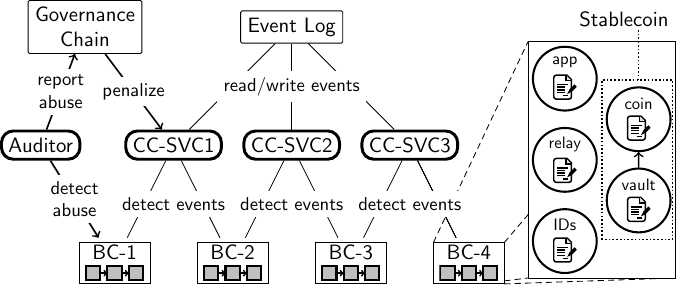}
\caption{Interaction between the main components of \blockx's core architecture.}
\label{fig:core_architecture}
\end{figure}

In this section, we describe \blockx's core architecture and its main components. 
\Cref{fig:core_architecture} visualizes the different entities and their interactions, and the different types of smart contracts on each chain.

\subsection{System Components}
\label{sec:system_components}

\subsubsection*{Blockchains (BCs)}
\blockx supports multiple independent blockchains that each support their own set of tokens, including the native token (e.g., Ethereum's ETH token) and user-created tokens (e.g., Ethereum's ERC-20 tokens and NFTs).
We assume that all blockchains  support smart contracts and use the account model for native token balances.\footnote{For blockchains that do not support smart contract and/or are UTXO-based, e.g., Bitcoin, we assume that a \textit{wrapped} version of its native token exists on a chain that meets our criteria \cite{lee2023sok}.} 

\subsubsection*{Cross-Chain Services (CC-SVCs)}
In \blockx, we use the \pie framework \cite{piechain} for communication between the underlying blockchains. In \pie, information is relayed between blockchains by CC-SVCs, which are entities that use full nodes or light clients to detect \textit{events} -- i.e., interactions with \blockx smart contracts. Each event type corresponds to a category of interactions of a similar nature, e.g., bids in a cross-chain auctions. Events that are detected by CC-SVCs are written to an \textit{event log}.
Users and CC-SVCs can subscribe to events and hence track the interactions with \blockx contracts across the supported blockchains. CC-SVCs earn a small fee for every trade that they facilitate -- this fee is processed by the smart contract for the \blockx application.

In \blockx{}, all nodes who participate in a deal must submit their tokens for escrow -- they are released after all nodes sign a commit vote or after the expiration of a timer. As such, even if CC-SVCs or the event log are compromised, then users are not at risk of having tokens stolen or frozen permanently. Although CC-SVCs can delay the conclusion of cross-chain deals, or misrepresent digital identities, they are disincentivized from doing so by \textit{auditors} (as discussed below).  As the same is true for the messaging service, we choose a performance-oriented solution, i.e., Apache's Kafka. The rarity of misbehavior in Ethereum's block proposal market \cite{heimbach2023ethereum}, which also relies on relay nodes to transmit information, empirically supports the assumption that relay services are sufficiently disincentivized by reputation damage in practice to refrain from misbehavior.

Instead of Kafka, a security-oriented solution such as a private blockchain -- e.g., Hyperledger Fabric \cite{cachin2016architecture} -- or Byzantine broadcast \cite{abraham2021good} could also be chosen for the messaging service. However, solutions that rely on Byzantine agreement such as private blockchains are known to have limited performance \cite{dinh2017blockbench}. In addition to its better performance, Kafka has a mature and well-maintained code base that would both be easier to deploy, have fewer bugs and exploits, and more rapidly patched in case of a bug than academic proposals and private blockchain platforms.

An example of a CC-SVC is the \textit{relayer} whose specification can be found in the \texttt{examples/ecomm/relayer} folder in \url{https://github.com/ntublockchain/I-O}. In particular, the code defines a set of Kafka events that are subscribed to (in this case, auction events such as incoming bids or the conclusion of the auction) and a set of listeners that specify the sort of blockchain behavior that prompt the CC-SVC to write new Kafka messages. We discuss the CC-SVC for the cross-chain marketplace example in more detail in \Cref{sec:decentralized_marketplace}.

\subsubsection*{Native Stablecoins}

Stablecoins are tokens whose value is pegged to a real-world asset, e.g., the US dollar. The use of stablecoins for cross-chain deals minimizes the influence of token price fluctuations on users' valuation of the involved assets. For example, if a user were to bid on an auction item using bitcoins, then a sudden change of the bitcoin price could cause the user to reconsider whether winning would lead to an acceptable outcome and abort. Although such a change of heart cannot be ruled out entirely as other offline circumstances may change (e.g., the user's valuation of the item), the use of stablecoins mitigates one prominent source of uncertainty.

To enable stablecoins in \blockx, we use the design of \croc \cite{crocodai} which relies on an optimized (to reduce volatility) portfolio of cryptoassets from multiple chains: customers who need stablecoins can buy them locally or deposit collateral tokens on supported chains. Collateral tokens are stored in dedicated smart contracts called \textit{vaults}, and can be reclaimed later at the cost of paying some interest. Stablecoin transfers are managed through \textit{coin} contracts -- \blockx contracts such as auctions call this contract to initiate transfers as they would for any other user-created token (e.g., ERC-20 tokens in Ethereum). If price changes or interest cause the ratio of the collateral's value to the amount of created stablecoins to become too low, then the collateral can be \textit{liquidated} through an auction. Price information about collateral tokens is provided to the vaults by price oracles (e.g., Chainlink or Uniswap contracts).   
Stablecoins can be transferred between chains if approved by the governance chain (see below). Governance nodes also decide on changes to the system-level parameters, e.g., the interest rate or liquidation ratio. To receive input from the governance layer, supported blockchains have \textit{relay} contracts that validate governance chain messages, e.g., by validating (group) signatures or a zero-knowledge proof-of-state.

Instead of a natively supported stablecoin such as \croc, users could choose to only trade existing stablecoins, e.g., Tether or Circle's USDCoin. The reasons to choose a natively supported stablecoin are twofold. First, it mitigates the risk to users from the sudden collapse of an external stablecoin, e.g., the collapse of the Terra/Luna stablecoin \cite{briola2023anatomy}. Although no stablecoin can be fully protected from price shocks and/or token collapses, \croc supports a diverse range of collateral and is therefore more resilient to catastrophic events that involve individual tokens \cite{crocodai}. The second that several components can be shared between \blockx and \croc: in particular, the same governance layer (as we discuss below) can be used to set system-level parameters such as fees, and CC-SVCs that transmit information between chains can also scan for cross-chain token transfer requests and relay them to the governance layer.

\subsubsection*{Digital Identities}

The pseudonyms of each user can be linked to a set of \textit{attributes} that represent important information about the user's identity, e.g., her name, location, or age. We assume that this information itself is stored (typically in encrypted form) on blockchains, e.g., in an \textit{ID contract} or using a dedicated data type. In \blockx, vendors can indicate during the specification of a cross-chain deal which attributes of potential customers must be submitted and what conditions must be met -- e.g., to ensure that customers submit (a hash of) their full name to prevent ticket scalping -- or they can provide aggregated feedback from users who meet certain attributes to prevent Sybil attacks. 
As part of any cross-chain deal, the creator must specify which ID ledgers are trusted, and which CC-SVCs are trusted to act as verifiers.
The end-to-end flow of obtaining and verifying a DID in \blockx is depicted in \Cref{fig:anoncreds} and discussed in more detail in \Cref{sec:digital_identities}.

\subsubsection*{Governance Chain}
The governance chain is a special blockchain (which can be an existing blockchain) on which nodes who monitor the underlying blockchains are present. The main role of governance chain nodes is to vote on changes to system-level parameters, approve cross-chain stablecoin transfers, and to verify claims of misbehavior by CC-SVCs. The governance chain can be an existing chain, e.g., Ethereum. In this case, \blockx governance would function analogously to a Decentralized Autonomous Organization (DAO \cite{sharma2023unpacking}) -- i.e., governance tokens would be tradeable, and major changes to the platform such as the addition/removal of CC-SVCs and changes to platform-wide parameters would be taken through majority votes. As part of every transaction facilitated by \blockx, a small fee is paid to the governance token holders, which gives them an incentive to participate beyond the ability to steer system-level decisions.

\subsubsection*{Auditors}

Misbehavior/abuse by CC-SVCs is detected by \blockx users called \textit{auditors}. Since the CC-SVCs' actions are entirely restricted to blockchain actions, misbehavior is provable to entities who have a view of the different blockchains. We identify three main types of misbehavior by CC-SVCs: i) not concluding a cross-chain deal fairly (e.g.,  not awarding an auction's item to the highest bidder), ii) causing a cross-chain deal to abort by failing to forward messages, and iii) misrepresenting the reputation or attributes of customers or vendors. Upon detecting misbehavior by a CC-SVC, an auditor can submit a claim of misbehavior to a smart contract on the governance chain. If the claim is invalid, then the auditor loses a small deposit, but if it is valid, the CC-SVC suffers a reputation penalty, which hampers the prospects of the CC-SVC being used in the future. 


\subsubsection*{Smart Contracts} As depicted in \Cref{fig:core_architecture}, each blockchain contains five main types of smart contracts: the coin and vault contracts for the stablecoin, the relay contract to receive input from the governance layer, and possibly an IDs contract to store DID information. Finally, a blockchain would also have one or more \textit{app} contracts to implement applications on top of \blockx -- in \Cref{sec:decentralized_marketplace}, we give an example of such an app, namely a cross-chain marketplace. 

\subsubsection*{Real-World Deployment of Components}

As mentioned previously, the prototype implementation on \url{https://github.com/ntublockchain/I-O} includes the code for the CC-SVCs and the integration with the \textit{coin} contract for the native stablecoin. As such, entities that intend to run CC-SVCs can re-use or extend this code. The implementation of auditors and the governance layer is left as future work, although the latter can be based on existing DAO smart contracts \cite{sharma2023unpacking}. Meanwhile, the code for the collateral vaults can be based on \croc or the existing Dai stablecoin.

The main incentive to operate an CC-SVC is to earn fees from processing trades. As such, they would likely attract a variety of entities depending on the size of the market supported by \blockx{} -- for example, 11 different relayers operated in Ethereum's block proposal market \cite{heimbach2023ethereum}, of which three were operated by the same entity (bloXroute). The incentives for auditors are similar to those of web certificate monitors in the Web PKI: this is done by parties who have an interest in the health of the ecosystem (e.g., other CC-SVCs) or aggrieved users.

\subsection{Digital Identifiers}
\label{sec:digital_identities}

\begin{figure}[tp!]
\centering
\includegraphics[width=\linewidth]{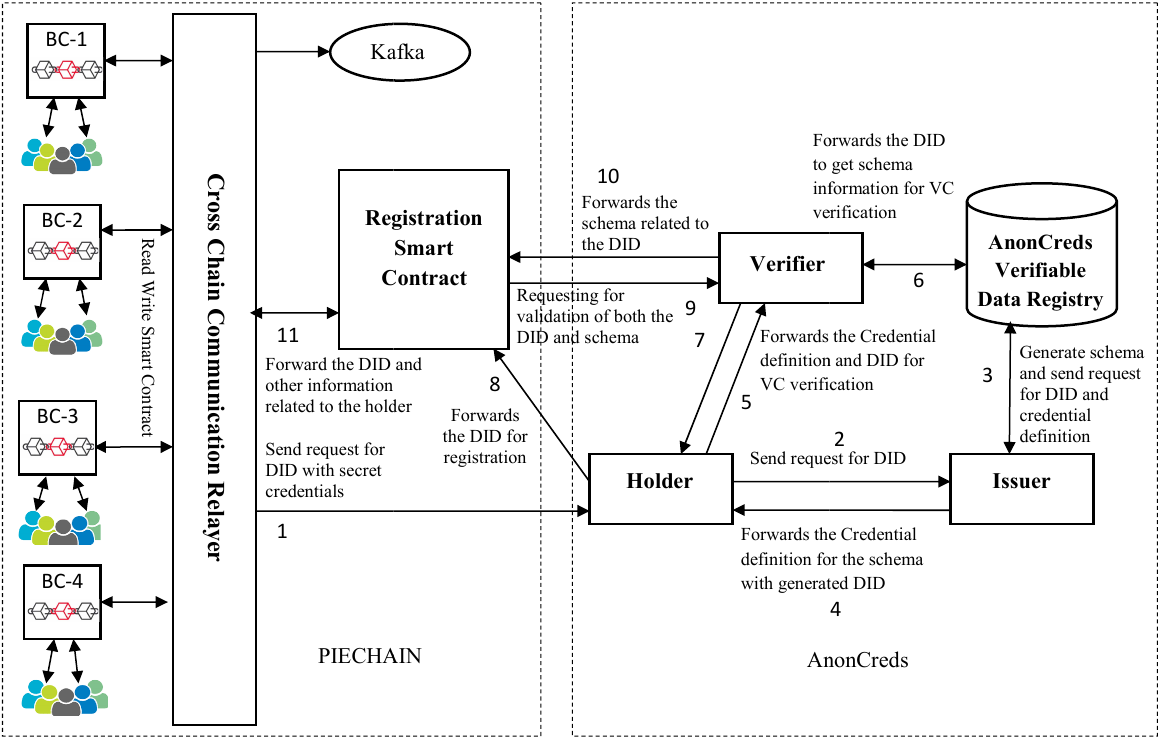}
\caption{\pie and AnonCreds integration \label{fig:anoncreds}} 
\end{figure}

The process of creating and verifying a DID is depicted in \Cref{fig:anoncreds}. We first discuss the entities in this figure, and then the steps of creation (1--4) and verification (5--11).

\subsubsection*{Entities}

In \blockx, a \textit{holder} can be any entity who tries to establish its credentials -- e.g., in the cross-chain marketplace, this could be a vendor or bidder. An \textit{issuer} can be any trusted organization that is specified by a participant in a trade.
 The \textit{VDR}, for which we use Hyperledger \mbox{AnonCreds} \cite{anoncreds}, stores the DIDs, schema information, credential definition identity, and revocation lists for future reference and the validation of VCs. 
\textit{Verifiers}, which in \blockx are CC-SVCs, communicate with the  Hyperledger AnonCreds  VDR \cite{anoncreds} and with the \textit{Registration Smart Contract} (RSC), which is a module of \pie that notifies verifiers through an event listing action. 

\subsubsection*{DID Creation}
Prior to the generation of DIDs, the issuer must first determine the schema and credential definition for the holder (which can be reused for many users/holders when suitable). 
Upon completion, both the issuer and verifier forward their requests to the system pool to obtain their identifiers, verification keys, and roles as Trust Anchor or Trustee respectively (step 1 of \Cref{fig:anoncreds}). After obtaining a suitable role (Trust Anchor), the holder asks the issuer (step 2) to request the AnonCreds VDR to generate the schema of the holder based on some secret credentials provided by the holder using a predefined reusable template (step 3). As a response, a schema ID is returned by AnonCreds VDR. Using this schema ID, the issuer requests the credential definition alongside other information such as tag values and type and revocation information for the schema.  Next, a DID is generated internally by the issuer. After getting credential definition information from AnonCreds, the issuer forwards it to the holder along with the DID (step 4). 

\subsubsection*{DID Verification}
To verify the DID, the holder first sends a verification request to the verifier (step 5). Next, the verifier obtains confirmation about the schema definition from AnonCreds (step 6), and either sends confirmation to the holder if the schema is valid, and rejects the request otherwise (step 7).
If successful, the holder sends the DID to the RSC (step 8), which emits a \pie event that prompts the verifier to validate the DID (step 9). Next, the verifier sends a confirmation to the RSC (step 10).
Finally, the successful verification of the DID is emitted as an event by the RSC (step 11), after which this information can be used by other \blockx smart contracts.

\subsection{Properties of Blockchain I/O}

We now summarize to what extent \blockx{} satisfies the desirable properties identified in \Cref{sec:initiatives}.

\subsubsection*{Primary Properties}
\blockx provides \textit{safety} because all involved parties  sign a commit vote before any exchange of assets, as such realizing a cross-chain deal as per Herlihy et al.\ \cite{herlihy2022cross}. In particular, CC-SVCs cannot misappropriate tokens as the aggrieved party would not sign the final commit vote. As in \cite{herlihy2022cross}, tokens are eventually released even if one party drops out or if the CC-SVC fails to transmit any of the messages. A single honest and online CC-SVC can provide liveness for any cross-chain trade, and CC-SVCs are disincentivized from going offline during a trade through the reputation system. If a CC-SVC still does go offline during a trade (e.g., an auction), then the trade can eventually be restarted, while the CC-SVC suffers a reputation penalty.

\subsubsection*{Secondary Properties}

The type of \textit{generality} supported by \blockx is that it can support any trade in which digital assets, including cryptocurrencies, are exchanged on multiple chains. Furthermore, \blockx is \textit{efficient} as the time costs of cross-chain interactions are in the order of seconds and gas costs are limited. We illustrate this through a case study involving a cross-chain marketplace in Sections~\ref{sec:decentralized_marketplace}~and~\ref{sec:experiments}.

\subsubsection*{Tertiary Properties}

\blockx achieves verifiability of user attributes through its AnonCreds component, which also achieves user \textit{privacy} through pseudonymity. \blockx satisfies \textit{price stability} through its \croc integration, and \textit{fairness} through the reputation system. If CC-SVCs violate fairness, e.g., by misrepresenting the winner of a cross-chain auction, or if they sign an incorrect ID proof, then this can be detected and penalized through the same reputation system as for liveness, which exploits a synergy between \blockx and the \pie and AnonCreds components.

\section{Cross-Chain Commerce: Use Case Examples}
\label{sec:use_cases}

In this section, we discuss three use cases for a decentralized e-commerce platform and discuss why the components discussed in the introduction provide non-trivial solutions to the challenges. 

\subsection{Cross-Chain Listings and Auctions}

As discussed in \Cref{sec:ecommerce}, the core application of e-commerce platforms such as eBay and Amazon is for users to offer items for sale either through a fixed-price listing or an auction. Several academic proposals have focused on guaranteeing privacy \cite{bag2019seal,blass2018strain} and fairness \cite{dreier2013defining} in this setting. However, price stability is also a requirement as sudden changes in token values may cause participants to regret and terminate their bids. Furthermore, as in regular e-commerce, a feedback and reputation mechanism allows customers to make informed purchase decisions. Listings and auctions are our core use case, and it will be the focus our implementation and experiments -- the remaining two case studies can be seen as extension of the first one.

\subsection{Scalping-Resistant Ticket Sales}


Ticket scalping refers to the practice where tickets are bought by third parties for the sole purpose of re-selling them at a higher price. Although ticket scalping may increase the efficiency of the sales process \cite{su2010optimal,bell2005ticket}, it is generally regarded as unfair by customers who observe tickets that were previously affordable being sold at prices that are (far) beyond their budget, and by vendors whose potential profits are usurped by a different entity \cite{bell2005ticket}.
Ticket scalping is non-trivial to avoid in a decentralized marketplace because the entities who make purchases are pseudonymous. For example, a scalper can trivially create a multitude of different accounts to circumvent restrictions on the number of tickets bought per user, and it is impossible by design to determine whether the customer who uses the ticket paid the original price or a higher price at an external marketplace. 


To address ticket scalping, we use the existence of dedicated blockchains that contain identity information to link ticket purchases to {verifiable identities}. These blockchains are typically different from the ones on which the ticket is sold and/or the payment is made, so this is necessarily a cross-chain challenge. In particular, as part of the function call that initiates the ticket purchase, the customer must also submit a DID. When the ticket is shown at the event, the customer reveals their name and/or any other associated information using an ID card to prove that it matches the DID, thus thwarting large-scale systematic ticket scalping.

\subsection{Sybil-Resistant Reputations}


As discussed in \Cref{sec:ecommerce}, reputation systems that allow users to rate their interactions with vendors may enhance their trust in the marketplace. One challenge in a reputation system is that a vendor may create Sybil accounts to boost its reputation or hurt its competitors' through dishonest feedback \cite{martins2020fostering,soska2016beaver}. Centralized systems can link each customer or vendor account to a credit card, making large-scale Sybil attacks impractical. However, in a fully decentralized anonymous marketplace, such an approach is impossible by design.




To address the challenge of Sybil attacks, we use verifiable identities to provide an analogous defense mechanism as a centralized marketplace. In particular, vendors who list an item can include a reputation metric signed by a cross-chain service, such that feedback is included in the metric only if it was issued by a customer who meets certain attributes, such as inclusion on a ledger maintained by trusted (e.g., government) organizations. Although this does not protect against Sybil attacks completely (e.g., a vendor could still ask family members or friends to give favorable feedback) it emulates the level of protection of centralized systems.

\section{Cross-Chain Marketplace}
\label{sec:decentralized_marketplace}
In this section, we present a proof-of-concept implementation of the first use case of \Cref{sec:use_cases}, a cross-chain marketplace, built on top of the \blockx infrastructure described in \Cref{sec:architecture}. We discuss both the core design of the marketplace and the smart contract implementation. 
We then explain how auditors detect abuse by CC-SVCs, and how to extend the implementation to the other use cases from \Cref{sec:use_cases}.

\subsection{Overview}

\subsubsection*{Users}

The marketplace has the following types of users.

\begin{itemize}
    \item \textit{Bidders} who hold digital (crypto) tokens and who are interested in purchasing listed tokens.
    \item \textit{Vendors} who want to sell tokens, and who seek to exchange them for (other) digital tokens.
\end{itemize}
In addition, the core user types of \blockx, i.e., CC-SVCs, auditors, and governance token holders, also participate in the marketplace.

\subsubsection*{Listings}

A listing represents an intent to sell a number of equivalent tokens -- these can be same-sized batches of cryptocurrencies, or NFTs that represent identical goods. 
Each listing belongs to one of the following common auction \cite{dreier2013defining} or listing types: 1) fixed-price listings, 2) open-bid increasing-price auctions, 3) closed-bid first-price auctions, 4) open-bid decreasing-price  auctions, and 5) closed-bid second-price auctions. Listings of type 1, 2, and 4 are resolved through three phases: bidding, conclusion, and feedback, whereas listings of type 3 and 5 are resolved through four phases: bidding, revealing, conclusion, and feedback. The actions in the four phases are as follows:

\begin{enumerate} 
\item \textit{Bidding.} Bidders submit their intent to purchase (fixed-price), their bid (open-bid auction), or their bid's hash (closed-bid auction), and transfer either the full bid or a minimum amount (the abort penalty) for escrow.
\item \textit{Revealing.} Users reveal their bids, and transfer the remaining value of their bid (i.e., their full bid minus the abort penalty) for escrow.
\item \textit{Conclusion.} A final transfer of assets is proposed, after which the  parties who transfer tokens vote to commit if agreeable. Upon commitment or abortion, the tokens are transferred or returned to the intended users. If an exchange is aborted because a user neglects to commit or reveal, then this user loses the abort penalty.
\item \textit{Feedback.} If the token represents physical items whose quality cannot be determined unambiguously, customers may provide feedback on the vendor -- e.g., to indicate their opinion of the speed of delivery, whether the item matched the description, etc.
\end{enumerate}

\subsubsection*{Events}

The following types of \pie{} events are recorded by the CC-SVCs:
\begin{inparaenum}
    \item \textit{AuctionCreationEvent}, emitted after a new listing has been created,
    \item \textit{BiddingAuctionEvent}, emitted after a new bid has been created, and
    \item \textit{AuctionEndingEvent}, emitted after a listing has been concluded.
\end{inparaenum}

\subsection{Smart Contract Functions}
\label{sec:marketplace_contract}
 Listings are processed through a single \textit{market} smart contract, which is a type of \textit{app} contract as depicted in \Cref{fig:core_architecture}. The \textit{market} contract has the following functions.

 \subsubsection*{createListing} Takes as input a start time, a reveal time, a conclusion time, a feedback time, CC-SVC addresses, a listing type, and an initial/fixed price. If successful, creates an entry for the listing using a hashmap in the \textit{market} contract. 

 \subsubsection*{bidFixed} Takes as input a listing ID. If the listing has the `fixed' type, the current time is between the start and conclusion times, and the sender has enough stablecoins in her wallet, then a bid is recorded, and an amount of stablecoins equal to the listing's fixed price is transferred to the \textit{market} contract for escrow.
 
 \subsubsection*{bidOpen} Takes as input a listing ID and bid value. If the listing has the `open' type, the current time is between the start and conclusion times, the bid value either exceeds the previous highest bid and the starting price (increasing-price auction) or is the first bid (decreasing-price auction), and the sender has enough stablecoins in her wallet, then a bid is recorded, and an amount of stablecoins equal to the bid value is transferred to the \textit{market} contract for escrow. 
 
 \subsubsection*{bidSealed} Takes as input a listing ID and bid hash. If the listing has the `sealed' type, the current time is between the start and reveal times, and the sender has enough stablecoins in her wallet to pay the abort fee, then a tentative bid is recorded, and an amount of stablecoins equal to the abort fee is transferred to the \textit{market} contract for escrow. 
 
 \subsubsection*{revealBid} Takes as input a listing ID and bid value. If the listing has the `sealed' type, the current time is between the reveal and conclusion times, the hash of the value equals the hash from \textit{bidSealed}, the sender has enough stablecoins in its wallet to pay the bid value minus the already escrowed abort fee, then a bid is recorded, and an amount of stablecoins equal to the bid value is transferred to the \textit{market} contract for escrow. 
 
 \subsubsection*{finAuction} Takes as input a listing ID, a number of winners, and a price (for second-price auctions). If called by the vendor, and if the current time is between the conclusion and feedback times, then the auction is concluded. Auctions that are not concluded before the feedback time are aborted and all tokens in escrow are returned.
 
 \subsubsection*{feedback} Takes as input a listing ID and a bid ID. If the auction has been concluded and the current time is between the conclusion and feedback times, then the user's feedback is recorded in a hashmap in the contract.

\subsection{Audits}

In the marketplace, CC-SVCs are relied on for faithfully concluding an auction through the \textit{concludeListing} function, and for signing DIDs for personal identifying information and for the feedback aggregates. If they misbehave in any of these roles, then this is detectable by auditors and governance token holders. For example, if they misrepresent the winning bidder, then a higher bid must exist on a chain than the one that was declared the winner. An auditor can send a proof of the existence of this bid to the governance chain, upon which a reputation penalty can be administrated to the CC-SVC. Similarly, if a CC-SVC has misrepresented a DID or reputation aggregate, then this can be demonstrated to the governance chain, as the zero-knowledge proof that was signed as valid by the CC-SVC must be invalid. This creates necessary checks and balances to ensure that end-users can identity and isolate misbehaving CC-SVCs, thus creating a mechanism to satisfy the implicit trust assumptions in \pie's design.

\subsection{Extension to Other Use Cases}
\label{sec:marketplace_use_cases}


The second and third use cases of \Cref{sec:use_cases} can be implemented by combining DIDs with the smart contract discussed in \Cref{sec:marketplace_contract} in the following way. To extend the auction use case to include ticket scalping protection, bids must only be accepted from users who have submitted a DID through a registration smart contract. During the creation of each auction, the auctioneer indicates which registration contracts are trusted. After purchasing a ticket through an auction, the user can reveal her personal information encapsulated in the DID upon redeeming the ticket.
Similarly, to extend the auction use case to offer Sybil-resistant reputations, users who provide feedback can also submit a DID from a registration contract. This allows other users to compute reputation metrics only from feedback supported by DIDs from trusted sources.



\section{Experiments}
\label{sec:experiments}
In this section, we present our experimental results to demonstrate that the marketplace built using \blockx (as discussed in \Cref{sec:decentralized_marketplace}) has practical performance. In particular, we evaluate the time and gas costs of the various steps of processing an NFT auction that accepts bids from multiple chains. We consider two main types of auctions: open-bid increasing (English) auctions and sealed-bid first-price auctions. The code for second-price and open-bid decreasing (Dutch) auctions can also be found in the online repository (\url{https://github.com/ntublockchain/I-O}), but the experimental results are similar to the other auction types and therefore not displayed here.
We perform two types of experiments: end-to-end experiments to illustrate the time costs of each step in the processing of an auction, and scalability experiments in which more than 2000 (2$\times$1152 bids in \Cref{tab:scalability_english_bid}) bids are placed on multiple auctions across all blockchains. 


\subsubsection*{Experimental Set-Up}

We use an iMac with an i9-10900k CPU to simulate an experimental environment with local test networks for Ethereum, Quorum, and Fabric. The NFT is implemented using an \textbf{Asset} on the  Fabric chain. For the stablecoin, we use a version of the Dai stablecoin for coin transfers based on the coin and relay contracts from \cite{crocodai}. In particular, we use the contract for coin transfers, but not for the creation of new stablecoins through collateral as this is orthogonal to our work. In our experiments, the different agent types all run on the same machine, so our results exclude network latency.


\subsection{End-To-End Experiments}

\subsubsection*{Steps}

We consider the following steps in the processing of a listing.

\textit{Create Listing.}
The auctioneer calls the \textit{addAsset} function of an \textbf{Asset} contract that is deployed on the Fabric network, which is detected by the CC-SVC. The CC-SVC deploys  \textbf{Auction} contracts on the Ethereum and Quorum platforms, and publishes an \textit{AuctionCreationEvent} to Kafka. 

\textit{Issue Bid.}
A bidder calls  \textit{bidSealed} on one of the deployed \textbf{Auction} contracts to issue a bid and submit stablecoins for escrow.
The CC-SVC detects a bid and successful coin transfer, and publishes a \textit{BiddingAuctionEvent} on Kafka. After the reveal time, the auctioneer and relayer end the bidding phase and start the reveal phase. Next, the bidders call \textit{revealBid} to reveal their bids. 

\textit{Conclude Listing.}
After the end of the conclusion timer, the auctioneer may conclude the listing by invoking a \textit{closeAuction} call to the \textbf{Asset} contract on Fabric. Alternatively, this action can be automatically triggered when the auction reaches its predetermined conclusion time.
Upon detection by the CC-SVC, it sends the listing's outcome to the \textbf{Auction} contract on each chain (i.e., whether the highest bid on that chain has won or not). This changes the state of these contracts to \textit{ending} -- which means that they await further action from the user -- and logs this activity as an \textit{AuctionClosingEvent}. 

\textit{Commit/Abort.}
The winning bidder either commits or aborts the auction result, which is detected and published on Kafka by the CC-SVC. 
The CC-SVC forwards the winner's response to the \textbf{Asset} contract on Fabric, then either returns or collects the coins transferred by the user in the previous stage.
When the related event \textbf{AuctionResponse} has been posted by one CC-SVC and received by Kafka, the CC-SVC transfers the asset from the auctioneer to the winner.

\textit{Feedback.}
If the auction result has been committed, then the winner can eventually submit her feedback about the purchased asset to the \textbf{Auction} contract on her chain.

\subsubsection*{Experimental Results}

An overview of the costs of steps mentioned above are displayed in \Cref{tab:process_listing}. For each step, we display the main entity responsible, the number of events, the gas costs on both blockchains (\textbf{bold}), and the time costs on the blockchains and Kafka (\textit{italic}). Deploying the auctions, initiating the reveal phase, and closing the auction are events that happen on both the Ethereum and Quorum blockchains, whereas adding the asset only happens on Fabric -- operations on Fabric do not have a gas cost. For the bids, we assume that there are 8 users on each of the two coin blockchains, leading to 16 bids, reveals, and withdrawals in total. For the bids, reveals, and withdrawals, we display the average costs over all events, and the half-width of an approximate 95\% confidence interval (based on the normal distribution) below the averages. The start auction, start reveal phase, and close auction events occur on two blockchains (Ethereum and Quorum), and the costs represent the sum across those blockchains. After the winner has been announced, only the winning bidder commits and gives feedback -- the losing bidders withdraw their bid from escrow, and the winner withdraws the asset. 

We observe reasonable time costs: each step, including those that occur on two blockchains, takes fewer than 6 seconds, whereas an auction would typically run for more than a day. Furthermore, a bid costs $\approx$115000 gas, which at current (early April 2024) gas prices ($\approx$30 GWei on average) would cost \$11.50 USD on Ethereum's main chain, but typically (much) less on other EVM-compliant chains (we emphasize that enabling bids on less-congested chains is a core motivation for \blockx). By summing all processing times, we observe that the entire auction (including all bid, reveal, and withdraw steps) can be concluded within 4 minutes.

\newcommand{\gaussconfintv}[2]{\begin{tabular}{c} #1 \\[-0.125cm] \scalebox{0.7}{$\pm$#2} \end{tabular}}

\setlength{\tabcolsep}{0.275em}

\begin{table}
\caption{Time costs in seconds (\textit{italic}) and gas costs (\textbf{bold}) of the various stages of processing a listing.}
\label{tab:process_listing}
\centering
\begin{tabular}{ccc|c|c|c|c|c|}
& & & \multicolumn{2}{c}{blockchains} & \multicolumn{1}{c}{Kafka} \\
entity &event & \# events & gas & time & time \\ \toprule
auctioneer & add asset & 1 & \textbf{0} & \textit{2.53} & \textit{0.013} \\ \midrule
relayer & start auction & 1 & \textbf{279664} & \textit{5.54} & \textit{0.012} \\ \midrule
bidder & bid & 16 & \gaussconfintv{\textbf{49243}}{10.4} & \gaussconfintv{\textit{1.06}}{0.275} & \gaussconfintv{\textit{0.018}}{0.002} \\ \midrule
auctioneer & end bidding phase & 1 & \textbf{0} & \textit{2.03} & \textit{0.002} \\ \midrule
relayer & start reveal phase & 1 & \textbf{59120} & \textit{2.12} & \textit{0.378} \\ \midrule
bidder & reveal bid & 16 & \gaussconfintv{\textbf{113778}}{2980.0} & \gaussconfintv{\textit{2.64}}{0.569} & \gaussconfintv{\textit{0.019}}{0.002} \\ \midrule
auctioneer & determine winner & 1 & \textbf{0} & \textit{2.08} & \textit{0.004} \\ \midrule
relayer & close auction & 1 & \textbf{66665} & \textit{5.52} & \textit{0.007} \\ \midrule
bidder & commit result & 1 & \textbf{63030} & \textit{0.906} & \textit{0.016} \\ \midrule
auctioneer & finalize auction & 1 & \textbf{29978} & \textit{5.1} & \textit{0.002} \\ \midrule
bidder & withdraw & 16 & \gaussconfintv{\textbf{30129}}{1083.0} & \gaussconfintv{\textit{1.18}}{0.32} & \gaussconfintv{\textit{0.016}}{0.002} \\ \midrule
bidder & feedback & 1 & \textbf{78745} & \textit{0.51} & \textit{0.017} \\
\bottomrule
\end{tabular}%
\end{table}

\subsection{Scalability}

\Cref{tab:scalability} displays experimental results in terms of processing time and gas costs for a setting with an increasing number of concurrent auctions and bids per auction. In particular, we consider a setting where 8 bidders each place $n$ bids on $n$ auctions that are simultaneously active, for $n=1,2,4,7$ (we also display $n=12$ for the open-bid auction). We focus on the bids and bid reveals, which, due to their frequency, have the most significant impact on the user experience. \Cref{tab:scalability_english_bid} displays the results for bids in an open-bid auction, \Cref{tab:scalability_sealed_bid} for bids in a sealed-bid auction, and \Cref{tab:scalability_sealed_reveal} for reveals in a sealed-bid auction.\footnote{We note that in a sealed-bid auction, a bidder would typically only bid once as there is no change in information during the bidding phase as the other bids are not revealed. However, the aim of this part of the experiments is to demonstrate that performance is not affected by a large number of bids.}  As in \Cref{tab:process_listing}, we display the averages over all events and the approximate 95\% confidence interval half-widths. However, unlike \Cref{tab:process_listing}, we separate the Ethereum and Quorum transactions -- e.g., note that the average 2.64 second processing time for bid reveals in \Cref{tab:process_listing} is the average of the 3.77 and 1.51 second time for Ethereum and Quorum, respectively.

\begin{table}
\caption{Gas costs (\textbf{bold}) and time costs in seconds (\textit{italic}) for an increasing number of auctions and bids per auction. Top: bids in an open-bid auction. Middle:  bids in a sealed-bid first-price auction. Bottom: reveals in a sealed-bid first-price auction.}
\label{tab:scalability}
\centering
\subfloat[][Open-Bid Auction (Bids)]{
\begin{tabular}{c|c|c|c|c|c|}
 \multicolumn{1}{c}{} & \multicolumn{1}{c}{} & \multicolumn{2}{c}{Ethereum} & \multicolumn{2}{c}{Quorum} \\ 
 \multicolumn{1}{c}{$n$} & \multicolumn{1}{c}{$\#$ bids} &\multicolumn{1}{c}{gas cost} & \multicolumn{1}{c}{time cost} &\multicolumn{1}{c}{gas cost} & \multicolumn{1}{c}{time cost} \\ \toprule

1 & 8 & \gaussconfintv{\textbf{114233}}{5502.0} & \gaussconfintv{\textit{3.84}}{0.02} & \gaussconfintv{\textbf{114233}}{5536.0} & \gaussconfintv{\textit{1.5}}{0.014} \\ \midrule

2 & 32 & \gaussconfintv{\textbf{101946}}{3280.0} & \gaussconfintv{\textit{3.94}}{0.15} & \gaussconfintv{\textbf{101946}}{3270.0} & \gaussconfintv{\textit{1.52}}{0.007} \\ \midrule

4 & 128 & \gaussconfintv{\textbf{98030}}{1046.0} & \gaussconfintv{\textit{3.98}}{0.072} & \gaussconfintv{\textbf{98030}}{1043.0} & \gaussconfintv{\textit{1.5}}{0.004} \\ \midrule

7 & 392 & \gaussconfintv{\textbf{96650}}{315.0} & \gaussconfintv{\textit{3.95}}{0.027} & \gaussconfintv{\textbf{96650}}{317.0} & \gaussconfintv{\textit{1.49}}{0.001} \\ \midrule

12 & 1152 & \gaussconfintv{\textbf{96457}}{286.0} & \gaussconfintv{\textit{3.95}}{0.016} & \gaussconfintv{\textbf{96623}}{166.0} & \gaussconfintv{\textit{1.49}}{0.001} \\ 

 \bottomrule
\end{tabular}%
\label{tab:scalability_english_bid}
}

\subfloat[][Sealed-Bid Auction (Bids)]{
\begin{tabular}{c|c|c|c|c|c|}
 \multicolumn{1}{c}{} & \multicolumn{1}{c}{} & \multicolumn{2}{c}{Ethereum} & \multicolumn{2}{c}{Quorum} \\ 
 \multicolumn{1}{c}{$n$} & \multicolumn{1}{c}{$\#$ bids} &\multicolumn{1}{c}{gas cost} & \multicolumn{1}{c}{time cost} &\multicolumn{1}{c}{gas cost} & \multicolumn{1}{c}{time cost} \\ \toprule

1 & 8 & \gaussconfintv{\textbf{49243}}{14.7} & \gaussconfintv{\textit{1.62}}{0.023} & \gaussconfintv{\textbf{49243}}{14.7} & \gaussconfintv{\textit{0.504}}{0.02} \\ \midrule

2 & 32 & \gaussconfintv{\textbf{41747}}{2597.0} & \gaussconfintv{\textit{2.04}}{0.099} & \gaussconfintv{\textbf{41747}}{2597.0} & \gaussconfintv{\textit{0.489}}{0.005} \\ \midrule

4 & 128 & \gaussconfintv{\textbf{37999}}{1125.0} & \gaussconfintv{\textit{1.94}}{0.052} & \gaussconfintv{\textbf{37999}}{1125.0} & \gaussconfintv{\textit{0.485}}{0.004} \\ \midrule

7 & 392 & \gaussconfintv{\textbf{36387}}{520.0} & \gaussconfintv{\textit{1.97}}{0.028} & \gaussconfintv{\textbf{36387}}{520.0} & \gaussconfintv{\textit{0.48}}{0.001} \\ 

 \bottomrule
 
\end{tabular}%
\label{tab:scalability_sealed_bid}
}

\subfloat[][Sealed-Bid Auction (Reveals)]{
\begin{tabular}{c|c|c|c|c|c|}
 \multicolumn{1}{c}{} & \multicolumn{1}{c}{} & \multicolumn{2}{c}{Ethereum} & \multicolumn{2}{c}{Quorum} \\ 
 \multicolumn{1}{c}{$n$} & \multicolumn{1}{c}{$\#$ reveals} &\multicolumn{1}{c}{gas cost} & \multicolumn{1}{c}{time cost} &\multicolumn{1}{c}{gas cost} & \multicolumn{1}{c}{time cost} \\ \toprule
 
1 & 8 & \gaussconfintv{\textbf{114716}}{5502.0} & \gaussconfintv{\textit{3.77}}{0.277} & \gaussconfintv{\textbf{112841}}{2098.0} & \gaussconfintv{\textit{1.51}}{0.017} \\ \midrule

2 & 16 & \gaussconfintv{\textbf{108153}}{5224.0} & \gaussconfintv{\textit{3.73}}{0.021} & \gaussconfintv{\textbf{107216}}{4226.0} & \gaussconfintv{\textit{1.51}}{0.011} \\ \midrule

4 & 32 & \gaussconfintv{\textbf{103934}}{3576.0} & \gaussconfintv{\textit{3.98}}{0.11} & \gaussconfintv{\textbf{103466}}{3152.0} & \gaussconfintv{\textit{1.49}}{0.004} \\ \midrule

7 & 56 & \gaussconfintv{\textbf{102126}}{2509.0} & \gaussconfintv{\textit{4.03}}{0.07} & \gaussconfintv{\textbf{101858}}{2319.0} & \gaussconfintv{\textit{1.49}}{0.003} \\ 
 \bottomrule
 
\end{tabular}%
\label{tab:scalability_sealed_reveal}
}
\end{table}

From \Cref{tab:scalability}, we observe that time costs in Quorum are typically lower than in Ethereum. Otherwise, we observe a strong correlation gas costs and processing times: higher gas costs imply higher processing times. We observe a mild decrease in gas or processing costs for all operations in \Cref{tab:scalability} -- the reason is that each user's first bid, and each auction's first reveal, are more costly than subsequent transactions because a new data point is added to a hashmap and updating this value in later transactions is cheaper than extending the hashmap. As the number of bids/reveals increases, the impact of this first transaction lessens, resulting in a decreasing average cost. Overall, we observe from \Cref{tab:scalability_english_bid} that our prototype implementation is able to handle over 2000 bids, which is far more than typical auctions which feature at most a few dozen bids \cite{balingit2009analysing,bodoh2021efficient}. If we sum all processing times, the 12 auctions for $n=12$ in \Cref{tab:scalability_english_bid} together take around 2.5 hours to complete, which, while exceptional, is still acceptable for auctions that may run for multiple days.

\section{Concluding Remarks}
\label{sec:conclusions}
We have presented \blockx, a framework for cross-chain commerce. It satisfies properties from three tiers of importance: primary (safety and liveness), secondary (efficiency and generality) and tertiary (pseudonymous digital identities, stablecoin support, and fairness). We have demonstrated the framework's versatility by creating a decentralized marketplace built on top of \blockx, hosting a variety of application use cases. We validated our proof-of-concept implementation for functional correctness and by benchmarking the overheads to demonstrate the practicality of the \blockx framework.

In future work, we hope to generalize \blockx to a wider range of use case, e.g., support for more advanced DeFi concept such as collateralized loans (with perhaps a credit rating stored as an attribute of a verifiable identity) or forms of insurance (such that the governance layer decides if certain bad events occurred). Future work may also explore the possibility of adding more tertiary properties to the set of desiderata for cross-chain commerce. Finally, an interesting direction for future work is to improve \blockx{}'s efficiency by aggregating bidder information such as DIDs and commit votes in a single zero-knowledge proof (e.g., a zk-SNARK or zk-STARK).


\section*{Acknowledgement} 
This work was supported by Ministry of Education (MOE) Singapore’s Tier 2 Grant Award No. MOE-T2EP20120-0003.

\bibliographystyle{plain}
\bibliography{references}

\end{document}